# Gamification of Quantum Mechanics Teaching

Ole Eggers Bjaelde
Centre for Science Education
Aarhus University
Denmark
oeb@phys.au.dk

Mads Kock Pedersen
Institute for Physics and Astronomy
Aarhus University
Denmark
madskock@phys.au.dk

Jacob Sherson
Institute for Physics and Astronomy
Aarhus University
Denmark
sherson@phys.au.dk

**Abstract**: In this small scale study we demonstrate how a gamified teaching setup can be used effectively to support student learning in a quantum mechanics course. The quantum mechanics games were research games, which were played during lectures and the learning was measured with a pretest/posttest method with promising results. The study works as a pilot study to guide the planning of quantum mechanics courses in the future at Aarhus University in Denmark.

## Introduction

Gameplay is becoming an increasing part of many aspects of our lives from socialisation to exercise - and higher education is no exception. Traditionally gameplay or gamification has been introduced in education to help students master the curriculum in untraditional ways to support learning and motivation for learning (Plass et al. 2013, Huang and Soman 2013, Kapp 2012). Here we further demonstrate how gamification can be used in higher education to include students in cutting-edge quantum mechanics research as well as to support learning and motivation for learning.

The present work is a small scale study of an intervention in a quantum mechanics course at Aarhus University, where traditional lectures were replaced by gaming sessions in which students would play a quantum mechanics game under guidance by the lecturer and teaching assistants. The outcome has been measured with a pretest and a posttest and indicates promising potential as a teaching method.

The games played in the intervention are developed as a citizen science project (Scienceathome 2014) in which gamers worldwide become important contributors to research on how to build a quantum computer (Weitenberg et al. 2011). The reason behind this is the experience from e.g. protein folding and neural mapping games such as Foldit (Cooper et al. 2010) and EteRNA (Lee et al. 2014) that humans sometimes have smarter problem-solving strategies than computers because of human's intrinsic intuition and ability to learn from mistakes. The games from the scienceathome project have earlier been used in gamified teaching interventions at the high school level (Magnussen et al. 2014).

## The Gamification Intervention

The gamification intervention was conducted in a second-year quantum mechanics course for all physics students at Aarhus University in Denmark. The number of students who participated in the intervention was 64. Before the intervention each student was asked 11 quantum mechanics questions in a pretest, using the clicker response technology (Vicens 2013). During the intervention students were given short instructions to the game and the content and were then challenged to perform concrete tasks in the game. Instructions and playing were subsequently alternating throughout the rest of the teaching session, which was two hours. At the end of the intervention, students were asked the same questions as in the pretest, again using clickers. In addition 11 evaluation questions were asked.

The content questions were chosen carefully to represent the content that was required in order to master the game. To give an example the first question was: How many nodes does the 5th eigenstate in a simple trapping potential have? With possible answers 4 (correct), 5 (wrong) and 6 (wrong).

11 games picked for the intervention represented different challenges in the research aiming at the development of a quantum computer, as well as illustrated specific subjects from the quantum mechanics course.

The games *ShakeIt* and *ShakeIt Up & Down* illustrate the problem of cooling down an excited atom using a classical analog of a ball rolling in a parabolic potential landscape. The intuition gained of the problems can then be applied in the quantum version of the problems *QShakeIt* and *QShake Up & Down*. The combined purpose of these exercises is to show that for some quantum research problems our classical world intuition is still applicable.

In contrast, we have chosen two games illustrating the process of quantum tunneling. It is a purely quantum phenomenon, in which a particle is able to cross a barrier even though it classically would be forbidden due to too little kinetic energy. It is a phenomenon which can be hard to gain an understanding of, since all prior knowledge usually contradicts the theory behind tunneling. In *Tunneling Expert* the players have to transport the atom through a small gap in a so-called zone of death (a region of the game window which causes the player to lose if the atom hits it). The gap is so small that the only possible solution is to tunnel the atom through the barrier. In *Bring Home Water Fast* the players have to pick up an atom from another potential well. From these exercises they learn some of the factors that affect the tunneling process such as the barrier height and the distance between the wells.

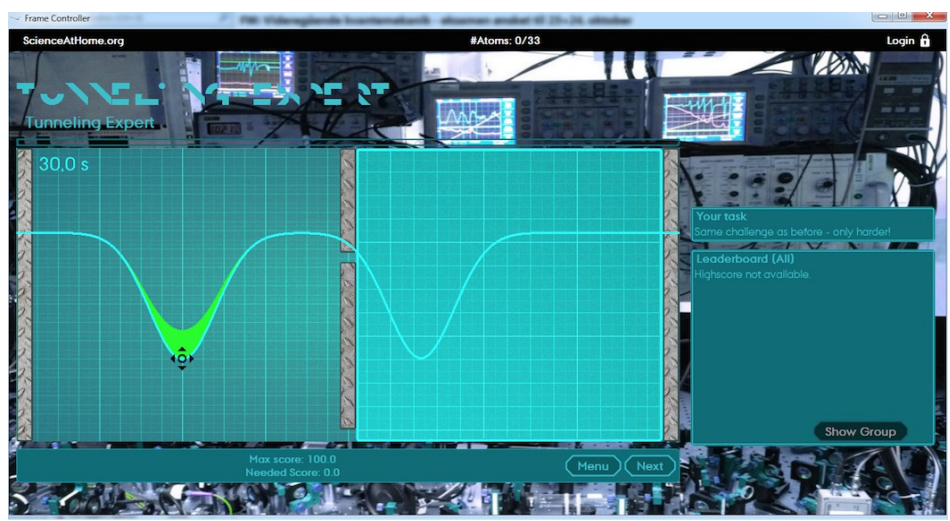

**Figure 1**: Screenshot from the *Tunneling Expert* game (Scienceathome 2014).

We also chose the five games that represent the scientific challenges we try to solve with the game. *Transport*, *Excited*, and *Merge* are all challenges where the solutions could help realize an actual quantum computer, and they aim at moving an atom to specific target states under different conditions. The two games *Wiggle Fast* and *Jaws!* are problems that current computer algorithms cannot solve, but where a study of how players solve the problems and their solutions could give insight into how new algorithms could be designed.

Besides the games the students were also given access to a simulation tool, which graphically simulated some of the basic text book problems usually met in an introductory quantum mechanics course e.g. the eigenstates of a particle in an infinite well.

The results from the pretest posttest experiment are shown in Fig. 2, where the blue vertical bars are pretest scores and red vertical bars are posttest scores. The students are grouped by their pretest performance; the missing blue bar to the left corresponds to students with zero correct answers in the pretest (1 student), the first blue bar students on the left students with one correct answer in the pretest (3 students) and so on. The red bar corresponds to the average score in the posttest for the same students, which for instance means that the student with zero correct answers in the pretest answered five questions correctly in the posttest. To measure how much the students learned during the intervention we define the gain according to the following equation:

$$\text{Gain} = \frac{\text{Correct answers in post-test - Correct answers in pre-test}}{\text{Number of questions (11) - Correct answers in pre-test}}$$

The gain captures how large a percentage of the number of questions the students answered incorrectly in the pretest, they now answered correctly in the posttest. If a student answered 5 questions correct in the pretest and answered 8 questions correct in the posttest it corresponds to a gain of 0.5. The gain is indicated by boxes in Fig. 2. Only students who answered more than 16 questions were included in Fig. 2 which is why the numbers below the vertical bars do not sum up to 64. The average gain across all students was 0.20, which is a very satisfactory number.

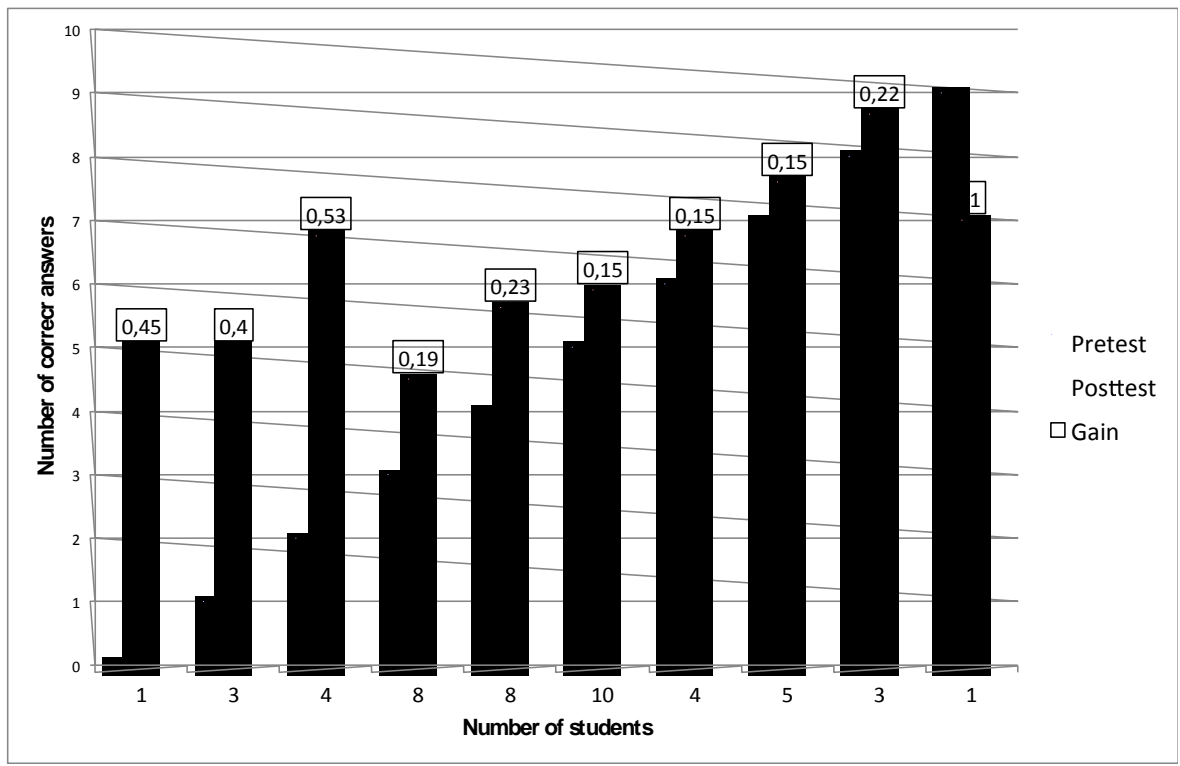

**Figure 2**: Pretest scores indicated by blue vertical bars, posttest scores indicated by red bars and collected gain indicated in boxes above the red bars.

Although the statistics is somewhat limited due to the small number of students (64) we can still draw some preliminary conclusions. First it seems that students who scored low on the pretest had a significant gain from the gameplay sessions. One could suspect that the reason for this gain was that these students were just randomly guessing. However, a closer look at data demonstrates that these students answered the same questions correctly in the pretest and posttest, which suggest that they are not just randomly guessing; they simply learned new content.

If we bin the data in Fig. 2 by the number of correct answers in the pretest, students with 0-3 correct in the pretest score a gain of 0.33, students with 4-6 correct in pretest score a gain of 0.18, while students with 7-9 correct in pretest score a gain of 0.05. This can partly be explained by diminishing returns (the higher the pretest score the harder it is to gain in the posttest), but could also be explained by the hypothesis that gamified education simply appeals to a different kind of students than more traditional teaching methods. More data is needed to support or dismiss this hypothesis. A similar result have earlier been observed by the PhET project where students who thought they knew the content of a simulation were less likely to explore the simulation and thus learned less from it (Adams et al. 2008).

In order to further investigate the collected gain for all students we did a cross-check with questions from the evaluation survey. Here we focus on three questions from the evaluation survey. 1) What was your final grade in the course Mechanics and Thermodynamics? 2) Did playing the game give you a better understanding of the content? 3) Was it fun to play the game?

The answers to these three questions are shown in the Fig. 3.

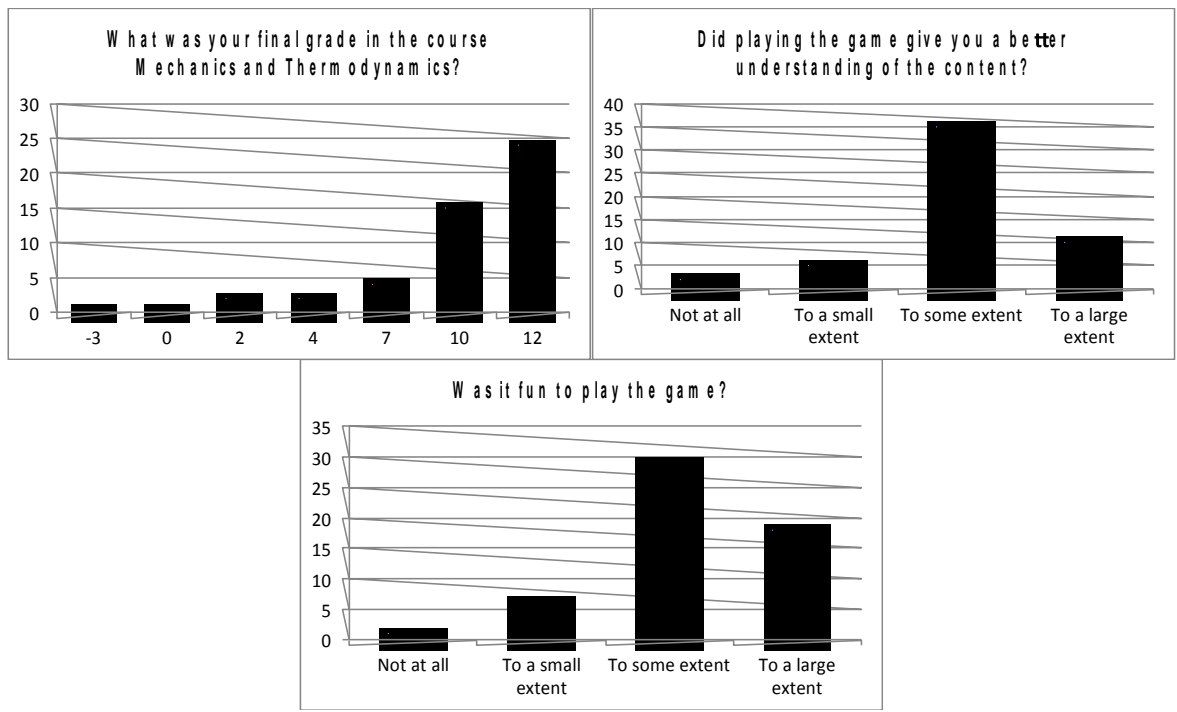

**Figure 3**: Distribtution of students' answers to three questions from the evaluation survey.

We now look at the gain for those students who scored the highest possible grade in the Mechanics and Thermodynamics course (Q#1 24 students), the students who indicated that they got a better understanding to a large extent (Q#2 10 students) and the students who answered that it was fun to a large extent (Q#3 18 students). The separate gains for these three groups is shown in Table.1.

| Question | Q#1 | Q#2 | Q#3 |
|---|---|---|---|
| Gain | 0.28 | 0.24 | 0.13 |

**Table 1**: The collected gain for three groups of students who scored the highest possible grade in the Mechanics and Thermodynamics course (Q#1 24 students), the students who indicated that they got a better understanding to a large extent (Q#2 10 students) and the students who answered that it was fun to a large extent (Q#3 18 students).

The most interesting result is that the students who had fun actually had a significantly lower gain (0.13) that the average student (0.20). This could be an indication that some students tend to become too focused on playing the game rather than the learning, which is certainly a challenge in gamified teaching setups (Adams et al. 2008). Secondly, we notice that students who felt they got a much better understanding of content also scored a higher than average gain (although not significantly) and thirdly, that the students who performed well in the Mechanics and Thermodynamics course scored a very high gain (0.28). This last result could indicate that good students respond well to gamified teaching - somewhat opposing the hypothesis presented after Fig.1. Because of the limited statistics we will defer from making any further conclusions on these numbers.

## Discussion

Gamification in higher education is a promising way of making students work with content in a structured way to support their learning. In addition, it is a good way of boosting the students' motivation for learning - especially when coupled to research as in our case (97 per cent of students in the present study responded that the research element gave them a higher or much higher motivation for learning).

We have demonstrated here in a small-scale study that on average students gain up to 20 per cent on posttest compared to pretest scores on quantum mechanics when exposed to a teaching session with gameplay. When we bin the numbers according to pretest scores, it becomes clear that the students who relatively learns the most are the students who scores lowest on the pretest. Interestingly, we also see a tendency for students with high grades learning a lot more than the average students. These two observations may seem to contradict each other, so further experiments are needed to draw any final conclusions here. Another point worth mentioning is the fact that students who responded to have had fun during the gameplay sessions actually performed significantly worse than the average student. It seems to indicate that some students get too focused thus hindering their learning.

The positive experiences from this intervention will help us reshape quantum mechanics courses taught at Aarhus University in the coming years with a focus on gamification elements.